\documentclass{article}

\usepackage{arxiv}

\usepackage[utf8]{inputenc} 
\usepackage[T1]{fontenc}    
\usepackage{hyperref}       
\usepackage{url}            
\usepackage{booktabs}       
\usepackage{amsfonts}       
\usepackage{nicefrac}       
\usepackage{microtype}      
\usepackage{xcolor}         

\usepackage[title]{appendix}
\usepackage{amsthm}
\usepackage{amsmath}
\usepackage{bbm}
\newtheorem{thm}{Theorem}
\newtheorem{corollary}{Corollary}

\usepackage{array}
\usepackage{todonotes}
\usepackage{comment}
\usepackage{csquotes}
\usepackage{array}
\newcolumntype{L}[1]{>{\raggedright\let\newline\\\arraybackslash\hspace{0pt}}m{#1}}

\usepackage{float}
\usepackage{amsmath}
\usepackage{mathtools}
\usepackage{algorithm2e}
\usepackage{physics}
\usepackage{subfigure}
\usepackage{xcolor}
\usepackage[export]{adjustbox}

\usepackage{tabularx}
\usepackage{multirow}

\newcommand{\Hi}{\mathcal{H}}

\newcommand{\R}{\mathbb{R}}
\newcommand{\Id}{\mathbb{I}}
\DeclareMathOperator*{\argmax}{arg\,max}
\DeclareMathOperator*{\argmin}{arg\,min}

\title{Large-Scale Quantum Separability Through a Reproducible Machine Learning Lens}

\author{%
  Balthazar Casalé
  \\
  Aix-Marseille University, CNRS, LIS \\
   Marseille, France \\
   \texttt{balthazar.casale@univ-amu.fr} \\
  \And
   Giuseppe Di Molfetta \\
   Aix-Marseille University, CNRS, LIS \\
   Marseille, France \\
   \texttt{giuseppe.di-molfetta@univ-amu.fr} \\
   \AND
   Sandrine Anthoine \\
   Aix-Marseille University, CNRS, I2M \\
   Marseille, France \\
   \texttt{sandrine.anthoine@univ-amu.fr} \\
   \And
   Hachem Kadri \\
   Aix-Marseille University, CNRS, LIS \\
   Marseille, France \\
   \texttt{hachem.kadri@univ-amu.fr} \\
}

\begin{document}

\maketitle

\begin{abstract}
  The \textit{quantum separability} problem consists in deciding whether a bipartite density matrix is entangled or separable. 
  In this work, we propose a machine learning pipeline for finding approximate solutions for this NP-hard problem in large-scale scenarios.
  We provide an efficient Frank-Wolfe-based algorithm to approximately seek the nearest separable density matrix and derive a systematic way for labeling density matrices as separable or entangled, allowing us to treat  quantum separability as a classification problem.
  Our method is applicable to any two-qudit mixed states.
  Numerical experiments with quantum states of 3- and 7-dimensional qudits validate the 
  {efficiency} of the proposed procedure, and demonstrate that it scales up to thousands of density matrices with a high quantum entanglement detection accuracy.
  This takes a step towards benchmarking quantum separability to support the development of more powerful entanglement detection techniques.

\end{abstract}

\section{Introduction}
\label{sec:intro}

Quantum entanglement is one of the most fundamental characteristics of quantum physics~\cite{horodecki2009quantum}. It is a key resource in quantum information theory and quantum computation. The description of entanglement is, however, very complex and its detection is particularly complicated. A central question is to determine whether a given quantum state is entangled or not~\cite{guhne2009entanglement,johnston2014entanglement}. Here we study this question  from a machine learning perspective.

A major difference of quantum computing and quantum information theory to their classical counterparts is that information is carried by qubits~\cite{nielsen_chuang_2010,rieffel2011quantum}.
Unlike bits which have only two possible states, 0 and 1, qubits can exist in those and any combination of them. More formally, a qubit is an element of a 2-dimensional Hilbert space $\mathcal{H}_2$, a complex inner product space that is also a complete metric space with respect to the distance function induced by the inner product. An arbitrary qubit $\ket{\psi}$\footnote{We use the Dirac's bra-ket notation. In this notation \enquote{bra} $\langle x |$ denotes a row vector and \enquote{ket} $|x\rangle$ a column vector.} 
may be written as $a_0\vert \psi_0 \rangle + a_1\vert \psi_1 \rangle$, where $\vert \psi_0 \rangle$ and $\vert \psi_1 \rangle$ form a complete orthonormal basis of $\mathcal{H}_2$.
For a single quantum state, we can also consider a Hilbert space with dimension larger than two. This gives rise to the concept of qudit, which is a generalization of the qubit to a $d$-dimensional Hilbert
space. For example, a qutrit ($d = 3$) is a three-state quantum system. 
A quantum system can also be divided into parts or subsystems. 
We focus here only on \textit{bipartite} quantum systems, i.e., systems composed of two distinct subsystems. The Hilbert space $\mathcal{H}$ associated with a {bipartite} quantum system is given by the tensor product $\mathcal{H}_A \otimes \mathcal{H}_B$ of the spaces $\mathcal{H}_A$ and $\mathcal{H}_B$ corresponding to each of the subsystems. 

A quantum state can be either \enquote{pure} or \enquote{mixed}.
While states of pure systems can be represented by a state vector $\psi \in \mathcal{H}$, for mixed states the characterization is done with density matrices $\rho$, {which are} positive semidefinite Hermitian matrices with trace equal to one.\footnote{Pure states can also be modeled with a density matrix allowing for uniform treatment, in this case $\rho = \psi\psi^\top$.}
The Hilbert space $\mathcal{H}$ associated with a {bipartite} quantum system is given by the tensor product $\mathcal{H}_A \otimes \mathcal{H}_B$ of the spaces $\mathcal{H}_A$ and $\mathcal{H}_B$ corresponding to each of the subsystems. 
This brings forward the notion of entanglement; any state that cannot be written as a tensor product of quantum states of the subsystems $A$ and $B$ is said to be entangled.

Quantum separability is the problem of deciding whether a \textit{bipartite} quantum state is entangled or not~\cite{ioannou2007computational,mintert2009basic}. If it is not entangled, the quantum state is said to be separable.
In this work, the quantum states are classically described via density matrices.
In the case of \textit{pure} states, this problem can be
efficiently solved using the Schmidt decomposition~\cite[Theorem 2.7]{nielsen_chuang_2010}. However, in the case of \textit{mixed} states, the problem is known to be NP-hard~\cite{gharibian2008strong, gurvits2003classical}.
It is still a serious challenge to efficiently find accurate approximate solutions to the quantum separability problem.
Recent works have therefore explored machine learning strategies to compute such approximations~\cite{asif2023entanglement,chen2021detecting,girardin2022building, lu2018separability}. 
 Although these works have made progress in using machine learning for quantum separability, there is still a notable gap between the two domains.
 In this work, we bridge this gap by providing a machine learning pipeline for quantum separability and entanglement detection in large-scale scenarios. Specifically, we make the following contributions:
\begin{itemize}
    \item we provide an efficient Frank-Wolfe-based algorithm to approximately seek the nearest separable bipartite density matrix;
    \item we derive a systematic way for labeling density matrices as separable or entangled, allowing us to treat  quantum separability as a classification problem;
    \item we propose a data augmentation strategy for entangled density matrices; 
    \item we create a new simulated labeled dataset with thousands of separable and entangled density matrices of sizes $9\times 9$ and $49 \times 49$; as far as we know, this is the first large-scale dataset for the quantum separability problem;
    \item  we perform numerical experiments using kernel SVM and neural networks to better understand the practical performance of nonlinear classifiers for quantum entanglement detection, providing  a reproducible baseline as well as identifying some limitations. 
\end{itemize}

    
\begin{table}[t]

		\label{tab:notation}
		\centering
		\def\arraystretch{1.5}%
		\begin{tabular}{lp{6cm}||p{1.1cm}L{4.5cm}}
$\mathcal{H}_A$ & Hilbert space of the  
quantum subsystem $A$ & $p_A$ & dimension of $\mathcal{H}_A$\\
$\mathcal{H}_B$ & Hilbert space of the  
quantum subsystem $B$ & $p_B$ & dimension of $\mathcal{H}_B$\\
$\mathcal{H}_A \otimes \mathcal{H}_B$ &  tensor product of  $\mathcal{H}_A$ and $\mathcal{H}_B$ & $p$ & dimension of $\mathcal{H}_A \otimes \mathcal{H}_B$ 
\\
 $\mathcal{S}_+^{p}$ & set of density matrices%
 &  $n$ & number of training data\\
 $\mathcal{S}_+^{p,\otimes}$ & set of separable density matrices%
 & $\mathcal{D}$ & training data $\{(\rho_i,y_i)\}_{i=1}^n$ \\
$M\geq 0$ & positive semi-definite~(psd) matrix & $\rho_i$ & input data in $\mathcal{S}_+^{p}$ \\
$M^\top$, $\|\cdot\|$  & transpose of $M$, Frobenius norm & $y_i$ & label in $\{-1,1\}$ \\
$\mathrm{tr}(M)$ & trace of $M$ & $y_i\!=\!-1$ & $\rho_i$ is separable\\
$M \otimes N$ & Kronecker product of matrices $M$ and $N$ & $y_i=1$ & $\rho_i$ is entangled \\
		\end{tabular}
	\def\arraystretch{1}%
 \medskip
		\caption{Notation summary.}
	\end{table} 
 

\paragraph{Reproducibility} We
release our code and data at \url{https://gitlab.lis-lab.fr/balthazar.casale/ML-Quant-Sep}. 
Our work takes a step towards benchmarking quantum separability and improving reproducibility.  We hope that our results will foster future computational works aimed at detecting quantum entanglement  in large-scale high-dimensional settings.

\paragraph{Notation} Table~\ref{tab:notation} summarizes the notation used in this paper. All Hilbert spaces are over the complex numbers.
The set of density matrices is $\mathcal{S}_+^{p} := \{\rho \in \mathbb{C}^{p\times p} : \rho \geq 0,  \mathrm{tr}(\rho)=1\}$, and
the set of separable density density matrices is $\mathcal{S}_+^{p,\otimes} := \{\rho \in \mathcal{S}_+^{p} : \rho=\sum_{j=1}^r q_j \sigma_{A,j} \otimes \sigma_{B,j}, \sigma_{A,j}\in \mathcal{S}_+^{p_A}, \sigma_{B,j}\in \mathcal{S}_+^{p_B}, q_j\in\mathbb{R}^+, \sum_j \!q_j\!=\!1\}$.

\section{A machine learning pipeline for quantum separability}
\label{sec:ml_qsep}

As mentioned in the introduction, there are some recent works trying to tackle the problem of quantum separability using machine learning (ML) techniques~\cite{asif2023entanglement,chen2021detecting,girardin2022building, lu2018separability,roik2021accuracy,urena2023entanglement}. However, all these previous studies are either limited to density matrices of very small dimensions, or require a lot of computations and are not feasible for large datasets.
The main difficulty stems from the fact that there is no well-labeled training data to learn from in this domain, especially for high-dimensional density matrices. This is the main problem that we want to solve.
The two main ingredients of our approach are:  \textit{positive partial transpose} (PPT) criterion and \textit{optimal entanglement witness}.

\subsection{Quantum separability}
\label{sec:qsep}
 The quantum separability problem for bipartite mixed quantum states is formally defined as follows:

\noindent\fbox{%
    \parbox{0.985\textwidth}{%
Let $\Hi := \Hi_A \otimes \Hi_B$ be a bipartite quantum state space and $\rho$ be a density matrix acting on $\Hi$. The quantum separability problem is to determine if $\rho\in {S}_+^{p}$ admits a decomposition of the form
\begin{equation}
\label{eq:quantum_sep}
   \rho = \sum_{j} q_j \; \sigma_{A, j} \otimes \sigma_{B,j} \;\;\; \text{s.t} \;\;\; q_j \in \R^+, \sum_{j} q_j = 1, 
\end{equation}
where $\sigma_{A,j}\in{S}_+^{p_A}$, $\sigma_{B,j}\in {S}_+^{p_B}$ are density matrices acting on $\Hi_A$ and $\Hi_B$, respectively, and the variables $\{q_j\}_j$ form a probability distribution.
\smallskip

If $\rho$ admits such a decomposition then it is said to be \textit{separable}, otherwise it is said to be \textit{entangled}. 
    }%
}

This problem is in general NP-hard~\cite{gurvits2003classical}.
Necessary and/or sufficient  conditions for quantum separability have been proposed; however, these conditions are rather restrictive or not computationally feasible (see~\cite{guhne2009entanglement, horodecki2009quantum} and references therein). 
In the following we present two criteria for quantum separability on which our approach is based.

\begin{figure}[t]
    \centering
    \includegraphics[scale=0.57]{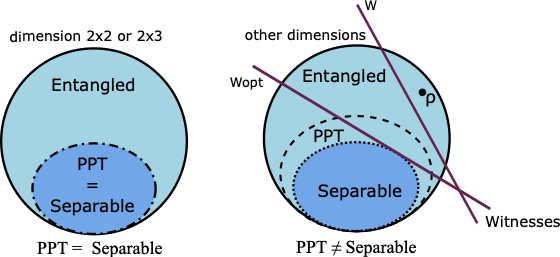}
    \caption{Schematic illustration of the set of separable, PPT and entangled bipartite quantum states. Left: When $p_A=2$ and $p_B=2$ or $p_B=3$, the positive partial trace (PPT) criterion is a necessary and sufficient condition for quantum separability, i.e., the set of separable states and the set of PPT states are the same. Right: In other dimensions, the set of separable states is included in the set of PPT states, i.e., all separable states are PPT but there exists PPT states which are not separable. If a state is not PPT, then it is entangled. The hyperplanes $W$ and $W_{opt}$ are entanglement witnesses; they detect the entanglement of the entangled density matrix $\rho$. The entanglement witness $W_{opt}$ defines a tangent hyperplane to the set of separable states, so it is optimal. }
    \label{fig:ppt}
\end{figure}

\paragraph{PPT criterion}

 Because of its simplicity, \textit{positive partial transpose} (PPT), also known as Peres-Horodecki criterion, is one of the most popular criteria for quantum separability~\cite{peres1996separability,horodecki1996necessary}.
Let $\rho$ be a density matrix on $\Hi$. The matrix $\rho$ is in $\mathcal{S}_+^p$ and then can be written as a {block matrix} composed of the matrix blocks $\rho_{i,j}$ of size $p_B \times p_B$, $\forall i,j=1,\ldots, p_A$. The {partial transpose} of $\rho$ with respect to subsystem $B$, denoted by $\rho^{\top_B}$, is defined as follows:
\[
\rho = \begin{bmatrix} \rho_{1,1} &   \cdots & \rho_{1,p_A} \\
                       \vdots &         &  \vdots        \\
                       \rho_{p_A,1} & \cdots &          \rho_{p_A,p_A} \end{bmatrix}
\Rightarrow
\rho^{\top_B} = \begin{bmatrix} \rho_{1,1}^\top &   \cdots & \rho_{1,p_A}^\top \\
                       \vdots &         &  \vdots        \\
                       \rho_{p_A,1}^\top & \cdots &          \rho_{p_A,p_A}^\top\end{bmatrix}.
\]
The PPT criterion states that if a density matrix $\rho$ is separable, then it has positive partial transpose, i.e., $\rho^{\top_B} \geq 0$. It is necessary for any separable density matrix to have positive partial transpose; yet in general this condition is not sufficient to guarantee separability, as there might be non-separable density matrices fulfilling the PPT condition. However this condition guarantees that if the partial transpose matrix has a negative eigenvalue, the state is entangled.
Moreover, if $\rho$ is a bipartite density matrix such that $p_A=2$ and $p_B=2$ or $p_B=3$, then $\rho^{\top_B} \geq 0$ implies that $\rho$ is separable. So the PPT criterion becomes in this case a necessary and sufficient condition~\cite{horodecki1996necessary}. In other dimensions this is not the case anymore~\cite{horodecki1997separability}. See Figure~\ref{fig:ppt} for a schematic illustration.

\paragraph{Entanglement witness}

An entanglement witness is a quantum separability criterion based on characterizing a separating hyperplane between the set of all separable density matrices and an entangled density matrix (see Figure~\ref{fig:ppt}).
Such a separating hyperplane always exists since the set of separable states  is convex~\cite{horodecki1996necessary}.
More formally, given an entangled density matrix $\rho\in\mathcal{S}_+^{p}$, 
an entanglement witness $W$ is a \textit{Hermitian} matrix in $\mathbb{C}^{p\times p}$ such that:
\begin{equation}
\label{eq:witness}
    \tr(W \rho_S) \geq 0 \text{ for all separable } \rho_S\in\mathcal{S}_+^{p,\otimes}, \text{ and } \tr(W\rho)<0.
\end{equation}
In this case we say that $W$ detects the entanglement of $\rho$, and then for each entangled density matrix there exists a
witness that detects its entanglement~\cite{horodecki1996necessary,guhne2009entanglement}.
In this work we are interested in the notion of \textit{optimal} entanglement witness~\cite{bertlmann2005optimal}. An entanglement witness $W_{opt}$ is optimal if there is a separable density matrix $\tilde{\rho}_S \in \mathcal{S}_+^{p,\otimes}$ such that $\langle W_{opt}, \tilde{\rho}_S \rangle=0$. The geometric interpretation of optimal entanglement witness is that it defines a tangent hyperplane to the set of separable density matrices, as illustrated in Figure~\ref{fig:ppt} (right).
One interesting feature of optimal entanglement witness is that it can be computed analytically and efficiently from the \textit{nearest separable} state  of an entangled density matrix $\rho$.
In other words, if $\tilde{\rho}_S$ is a separable density matrix that satisfies $\|\tilde{\rho}_S - \rho\| = \min_{\rho_S\in\mathcal{S}_+^{p,\otimes}} \|\rho_S - \rho\|$, where $\|\cdot\|$ is the Frobenius norm, then 
\begin{equation}
\label{eq:optimal_witness}
    W_{\rho,opt} = \frac{\tilde{\rho}_S - \rho - \langle \tilde{\rho}_S, \tilde{\rho}_S - \rho\rangle \Id_p}{\|\tilde{\rho}_S - \rho\|},
\end{equation}
where $\Id_p$ is the identity matrix, is an optimal entanglement witness~\cite{bertlmann2005optimal}. It is easy to check that $\langle W_{\rho,opt}, \tilde{\rho}_S \rangle=0$.

\subsection{Proposed strategy}

Most of the previous works on ML-based quantum separability have considered situations where the dimension of the subsystem $A$ is equal to 2 ($p_A=2$) and the dimension of the subsystem $B$ is equal to 2 or 3 ($p_B=2$ or $p_B=3$), and used the PPT criterion to label the training density matrix as separable or entangled~\cite{asif2023entanglement,chen2021detecting,roik2021accuracy,urena2023entanglement}. But this leads to a partial view of the problem, since the learning algorithm in this case should represent and learn whether a quantum state is PPT or not; this is different from the task of quantum separability in general and is not really useful. So, the critical point is the lack of prior knowledge on density matrices that are both PPT and entangled to be incorporated during the learning process. 
We tackle this problem and propose a quantum entanglement detection strategy that scales well to high-dimensional density matrices.

It is worth noting that the PPT criterion could be used to label density matrices that are entangled and not PPT. As discussed in Section~\ref{sec:qsep}, quantum states that are not PPT are entangled.
Our solution to generate density matrices that are PPT and entangled is based on the notion of entanglement witness. More precisely, we first provide an efficient Frank-Wolfe algorithm to compute good approximations of the nearest separable state problem in order to identify optimal entanglement witnesses. This will allow us to detect the entanglement of PPT density matrices that are not separable. We then propose a data augmentation strategy for these entangled density matrices. 

\subsubsection{A Frank-Wolfe algorithm to construct nearest separable approximations}

Finding the nearest separable state to a given density matrix is a {constrained convex optimisation problem}.
Let $\rho\in\mathcal{S}_+^p$ be a density matrix, the nearest separable state $\tilde{\rho}$ is the solution of the following minimization problem
\begin{equation}
\label{eq:nearest_sep}
  \tilde{\rho} := \argmin_{\sigma \in \mathcal{S}_+^{p,\otimes}} f_\rho(\sigma),
\end{equation}
where $f_\rho(\sigma) := \|\sigma - \rho\|^2$ {with $\| .\|$ the Frobenius norm} and $\mathcal{S}_+^{p,\otimes}$ is the set of separable density matrices.

In the field of machine learning, the Frank-Wolfe (FW) algorithm has received a considerable interest due its \textit{projection-free} property~\cite{jaggi2013revisiting}. This leads to efficient algorithms capable of handling large datasets. A pseudo code of the FW algorithm is outlined in Algorithm~\ref{alg:FW}.
It is worth to mention that this method was applied in~\cite{dahl2007tensor} to find the nearest separable density matrix. It was shown that this problem has a unique optimal solution and that the linear subproblem~(line 3 in Algorithm~\ref{alg:FW}) leads in this case to the following optimization problem:
\begin{equation}
\label{eq:subproblem}
    \argmax_{\ket{s_A}\in\mathcal{H}_A, \ket{s_B}\in\mathcal{H}_B} \Big\langle (\rho - \sigma_t) \ket{s_A}\otimes\ket{s_B}, \ket{s_A}\otimes \ket{s_B} \Big\rangle.
\end{equation}
A block coordinate ascent method has been applied to this problem, requiring an iterative procedure and alternating optimization over $\ket{s_A}$ and $\ket{s_B}$.
The whole algorithm becomes prohibitively expensive and  infeasible even for small values of $p_A$ and $p_B$, the dimensions of $\mathcal{H}_A$ and $\mathcal{H}_B$.  

We provide a computationally efficient variant of this Frank-Wolfe method. We propose a simple procedure to approximate the solution of the linear subproblem~\eqref{eq:subproblem}.
Our strategy consists of two steps.
First, we compute 
    $\ket{s^*} = \arg\max_{\ket{s} \in \Hi} \Big\langle (\rho - \sigma_t) \ket{s}, \ket{s}\Big\rangle$,
    which amounts to finding the normalized eigenvector associated to the largest eigenvalue of the matrix $(\rho - \sigma_t)$. 
    This corresponds to solving the optimization problem~\eqref{eq:subproblem} without considering the constraints, i.e., $\ket{s^*}$ may not necessary be a tensor product of two states in $\mathcal{H}_A$ and $\mathcal{H}_B$.
    This step can be done efficiently using the power method.
    Then, we construct the matrix $\sigma^*$ from the Schmidt decomposition of the state $\ket{s^*}$. Let  $\ket{s^*} = \sum_{i} \lambda_i \ket{a_{i}} \otimes \ket{b_{i}}$ be the Schmidt decomposition of $\ket{s^*}$ (see Appendix~\ref{sec:appendix_schmidt} for more details), then
    $\sigma^* := \ketbra{a_1}{a_1} \otimes \ketbra{b_1}{b_1}$, where $\ket{a_1}$ and $\ket{b_1}$ are the quantum states  corresponding to the largest Schmidt coefficient $\lambda_1$. 
The Schmidt decomposition allows us to compute the best separable approximation of a pure state, i.e., $\ket{a_1}, \ket{b_1}=\argmin_{\ket{a},\ket{b}} \|\ket{s^*} - \ket{a} \otimes\ket{b}\|$.
    Note that there is no need to compute all the states of the decomposition, only the two corresponding to $\lambda_1$ are required. This has the same cost as computing the first pair of left and right singular vectors.
Our Frank-Wolfe algorithm for approximating the nearest separable density matrix is depicted in Algorithm~\ref{alg:QFW}.

\begin{figure}[!t]
\begin{minipage}{0.45\textwidth}
\begin{algorithm}[H]
\LinesNumbered
\KwData{$f : D \mapsto \R$ convex, differentiable function \newline
        $S \subset D$ convex set \newline}
\KwResult{${\sigma}^* \in S \approx \argmin_{\sigma \in S} f(\sigma)$ \newline}

$\sigma_0 \gets \sigma \in S$ \;
\smallskip
\For{$t = 0$ \KwTo $T$}{
    $\sigma^* \gets \argmin_{\sigma \in S} \sigma^\top \nabla f(\sigma_t)$ \;
   \smallskip
    $\alpha \gets \frac{2}{(t+2)}$ \;
    \smallskip
    $\sigma_{t+1} \gets (1 - \alpha) \cdot \sigma_t + \alpha \cdot \sigma^*$ \;
}
\smallskip
\Return $\sigma_T$ \;
\medskip
\caption{FW algorithm.}
\label{alg:FW}
\end{algorithm}
\end{minipage}
\begin{minipage}{0.55\textwidth}
\begin{algorithm}[H]
\KwData{$\rho \in \mathcal{S}_+^{p}$ a density matrix\newline}
\KwResult{$\rho^* \in \mathcal{S}_+^{p,\otimes}$ a separable density matrix approximating~\eqref{eq:nearest_sep}\newline}

$\rho_0 \gets  \ketbra{\psi_A}{\psi_A} \!\otimes\! \ketbra{\psi_B}{\psi_B}$ (random state)\;
\smallskip
\For{$i \gets 0$ \KwTo $T$}{
    $\ket{s^*} \gets \text{ largest eigenvector of }(\rho - \rho_t)$ \;
    \smallskip
    $\ket{s_A^*}, \ket{s_B^*} \gets \text{largest  Schmidt states\footnotemark of } \ket{s^*}$ \;
    \smallskip
    $\alpha \gets \frac{2}{t+2}$\;
    \smallskip
    $\rho_{t+1} \gets (1 - \alpha)  \rho_t + \alpha \ketbra{s_A^*}{s_A^*} \otimes \ketbra{s_B^*}{s_B^*}$ \;
}
\smallskip
\Return $\rho_T $ \;
\medskip
\caption{Nearest separable density matrix with FW algorithm and approximate linear subproblem.}
\label{alg:QFW}
\end{algorithm}
\end{minipage}
\end{figure}
\footnotetext{The states corresponding to the largest Schmidt coefficient of a Schmidt decomposition.}

\begin{figure}
  \begin{minipage}{0.53\textwidth}
    \centering
    \includegraphics[width=0.75\linewidth]{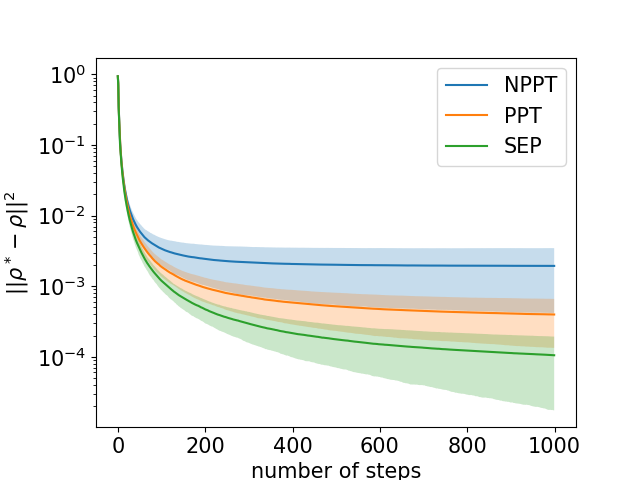}
  \end{minipage}%
 \hspace*{-1cm} \begin{minipage}{0.53\textwidth}
    \centering
    \includegraphics[width=0.75\linewidth]{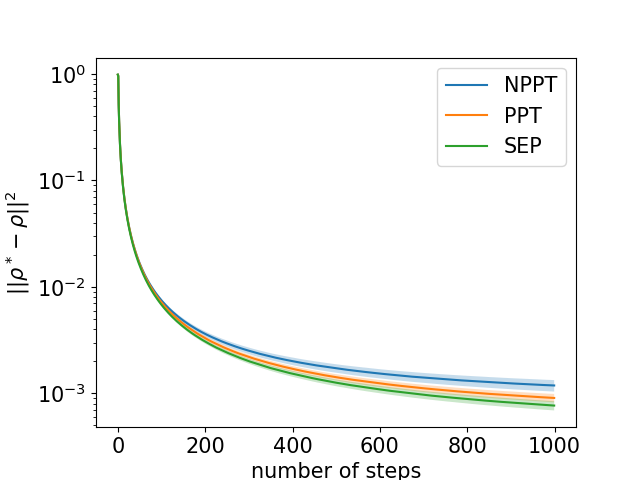}
  \end{minipage}
  \caption{Evolution of the approximation error of the proposed FW algorithm (Algorithm~\ref{alg:QFW}) over iterations for randomly generated density matrices which are Non-PPT (blue), PPT (orange) and separable (green) in state spaces of dimensions $p_A = p_B = 3$ (left) and $p_A = p_B = 7$ (right). The colored area around each curve corresponds to the standard deviation. One-thousand data samples were generated for each class of density matrices.}
\label{fig:dist_by_step}
\end{figure}

Figure~\ref{fig:dist_by_step} shows the approximation error of Algorithm~\ref{alg:QFW} over iterations in the cases of separable, PPT and non-PPT quantum states. The error is smaller for separable states. This is expected since when $\rho$ is separable, the FW algorithm should converge to this state. Note that PPT states contain separable and entangled states, while non-PPT states are all entangled. 
Results in Figure~\ref{fig:dist_by_step} also show that our FW algorithm can be used for detecting entanglement: the distance $\|\rho^*-\rho\|$ is compared to a threshold and an assignment to the label \enquote{separable} or \enquote{entangled} is made based
upon whether the distance is below or above the threshold, respectively. 
There are two main issues with this strategy: (i) we do not know how to choose the threshold, and (ii) we need to run the FW algorithm on all the test data points, which renders inference costly. Therefore, we instead train a classifier to
\\[0.1cm]
\centerline{\textit{learn from data the decision function between separable and entangled states}.} \\[0.1cm]
As we show in the following section, the proposed efficient FW algorithm is the key to create \enquote{well-labeled} training datasets for this task.

\subsection{Density matrix data labelling and augmentation}
\label{sec:label_augm}

\begin{figure}[t]
\centering
\includegraphics[scale=0.25,valign=b]{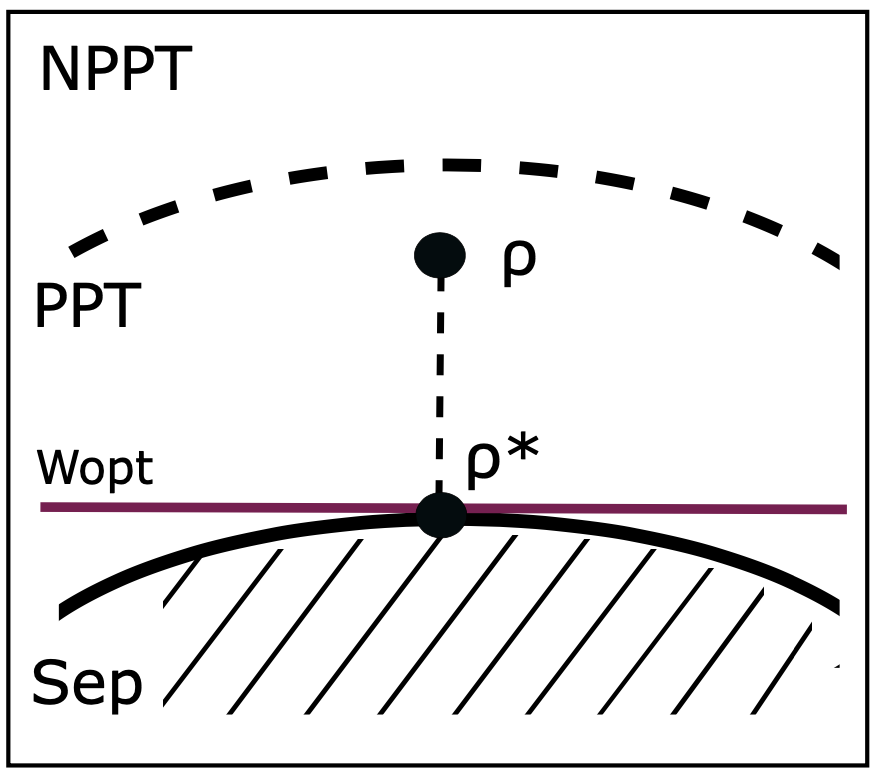}
$\qquad\qquad\qquad$ \includegraphics[trim={-2cm -0.04cm 0 0},scale=0.25,valign=b]{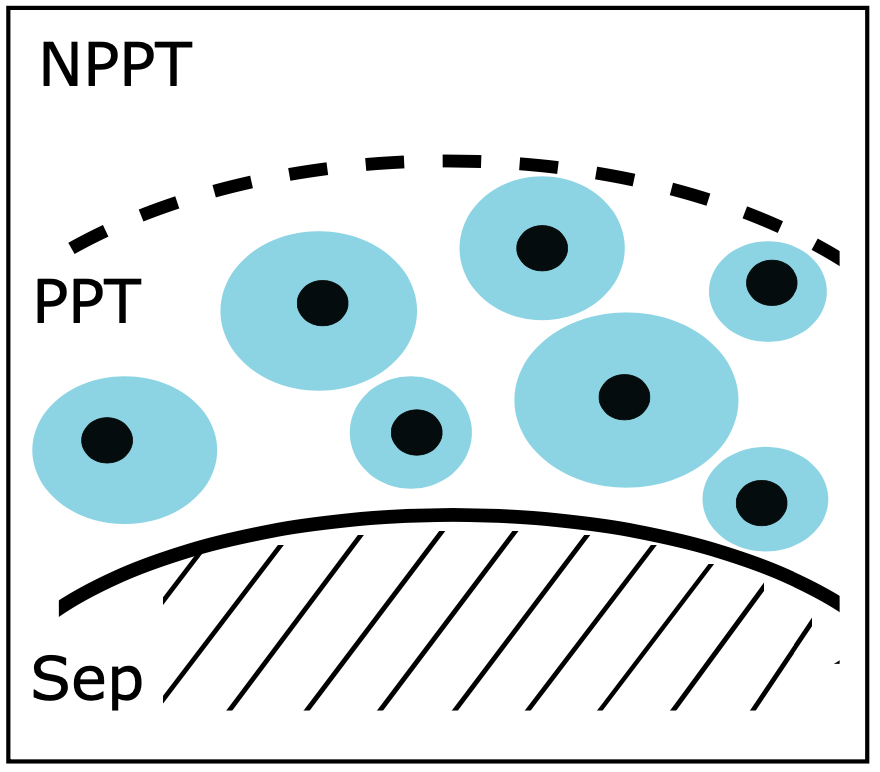}
\caption{A schematic depiction of the density matrix data labelling and augmentation process. Left: For each generated PPT density matrix $\rho$, we approximate its nearest separable quantum state $\rho^*$ using the FW method (Algorithm~\ref{alg:QFW}). We then compute its optimal entanglement witness $W_{opt}$ using~\eqref{eq:optimal_witness} and check numerically whether it satisfies the assumptions in~\eqref{eq:witness}. If so, then it is a valid witness and $\rho$ is an entangled PPT state. Right: For each entangled PPT state, we identify using the computed optimal entanglement witness a region that includes this state and preserves inseparability (i.e., entanglement) and positivity under small perturbations. We randomly generate density matrices in these regions. All these density matrices are PPT entangled states.}
\label{fig:data_labelling_augmentation}
\end{figure}

Our quantum separability data generation model is based on two steps: a density matrix labelling process and a data augmentation scheme (see Figure~\ref{fig:data_labelling_augmentation}).  

\paragraph{Creating a labeled quantum separability dataset} Our dataset contains three types of bipartite density matrices: separable (denoted by SEP), PPT entangled (denoted PPT-ENT) and non-PPT~(denoted NPTT-ENT). Recall that by the PPT criterion, if a quantum state is non-PPT then it is entangled (see Section~\ref{sec:qsep}).
The dataset is based on random density matrices. 
For the generation of random  density matrices, we refer the reader to~\cite{zyczkowski2011generating,aubrun2014random, aubrun2014entanglement}.
Data for class SEP are easy to generate. One can randomly generate  density matrices of sizes $p_A \times p_A$ and $p_B \times p_B$ and use~\eqref{eq:quantum_sep} to compute a separable density matrix in $\mathcal{S}_+^{p,\otimes}$, so the separability here holds by construction. 
Data for class NPTT-ENT are also easy to obtain. We sample random density matrices in $\mathcal{S}_+^{p}$ and use the PPT criterion to identify non-PPT quantum states.
The task of constructing data for quantum density matrices that are both PPT and entangled is more challenging.

To address this challenge, we proceed as follows.
We generate random density matrices in $\mathcal{S}_+^{p}$ and apply the PPT criterion to select only PPT states. At this stage, the obtained density matrices can be separable or entangled. The goal now is to identify those who are entangled. To do so we compute for each PPT density matrix the nearest separable quantum state using Algorithm~\ref{alg:QFW}, with the objective to determine their optimal entanglement witnesses as given by~\eqref{eq:optimal_witness}. If the witness $W_{opt}$ for a PPT density matrix $\rho$ is valid, i.e., $\tr(W_{opt} \rho_S) \geq 0$ for all separable $\rho_S\in\mathcal{S}_+^{p,\otimes}$ and $\tr(W_{opt}\rho)<0$ (see Eq.~\ref{eq:witness}), then $\rho$ is entangled.
We numerically check these assumptions by generating ten thousands separable density matrices and selecting only the PPT states that satisfy the assumptions on these generated separable states. By doing so, the chosen PPT density matrices are very likely to be entangled. To the best of our knowledge, this is the first practical and efficient method that can provide general PPT entangled density matrices. 
Note that some examples of PPT entangled states can be found in the literature, see, e.g.,~\cite{halder2019family,sindici2018simple}. However, they are very specific and restricted to particular configurations~(specific dimension requirements, for example), which renders them unsuitable for learning a good decision rule.

\paragraph{Data augmentation} The procedure for generating PPT entangled density matrices can be computationally expensive, especially when the dimension $p$ grows. To alleviate this issue, we propose a density matrix data augmentation scheme inspired by a work on robustness of bipartite entangled states~\cite{bandyopadhyay2008robustness}.
The idea is to add noise to PPT entangled states in order to generate new states while preserving  inseparability and positivity under PPT operations.
Let $\rho \in \mathcal{S}_{+}^p$ be a PPT entangled density matrix. There is a region containing $\rho$ inside of which all the density matrices are also PPT and entangled~\cite[Theorem 1]{bandyopadhyay2008robustness}.
The variables needed to define this region can all be computed in closed-form from an entanglement witness $W$ that detects the entanglement of $\rho$. The complete analytic characterization of this region is given in Appendix~\ref{sec:appendix_robust}.

During the data labelling stage, a number of PPT entangled density matrices are generated using  optimal entanglement witnesses. These witnesses will be used to characterize different regions containing only PPT entangled states. Adding noise to the initial density matrices while remaining in these regions produces new PPT entangled states (See Figure~\ref{fig:data_labelling_augmentation}).
This procedure allows to generate an arbitrarily large number of PPT entangled density matrices from a few  examples; however, if the region is small, all the augmented examples will be similar. To add more variability in the data augmentation process, we also perform random transformations of the form $U \rho U^{\dagger}$, where the symbol~${\dagger}$ denotes the conjugate transpose, $U := U_A \otimes U_B$,   and $U_A, U_B$ are random unitary matrices acting on $\Hi_A$ and $\Hi_B$, respectively. This is a separable transformation that does not change the entanglement. Indeed, it is easy to see that $U \rho U^{\dagger}$ is a separable density matrix when $\rho$ is separable.

\section{Related work}

The majority of the literature on ML-based bipartite quantum separability has focused only on the case where the quantum subsystems are of dimensions $2 \times 2$ and $2 \times 3$ (see, e.g.,~\cite{asif2023entanglement,chen2021detecting,urena2023entanglement,roik2021accuracy}).  The PPT criterion can be used in this case to label the training density matrix as separable or entangled. But this leads to a partial view of the problem, since the learning algorithm in this case should represent and learn whether a quantum state is PPT or not; this is different from the task of quantum separability in general and is not really useful. So, the critical point is the lack of prior knowledge on density matrices that are both PPT and entangled to be incorporated during the learning process.

The closest works to ours are~\cite{lu2018separability,girardin2022building}. In~\cite{lu2018separability}, a convex hull approximation (CHA) algorithm is proposed to approximate the set of separable states. This is computationally heavy since a large number of random separable states is needed to well approximate the convex set. A bagging approach is used to reduce the computational cost. However, CHA must be applied for all training and test data, which renders the method computationally infeasible for application to moderate- or large-scale density matrices.
The only work we are aware of that analyzed learning scenarios for quantum separability with high dimensional quantum states is the recent paper~\cite{girardin2022building}. Although they did not provide a mechanism for entangled PPT data generation and augmentation, which is central to our task, the authors proposed a generative neural network to find a closest separable state to a given target quantum density matrix which works up to dimension $10 \times 10$.

\section{Experiments}

We created two fully labeled datasets of 6,000 bipartite density matrices following the process described in Section~\ref{sec:ml_qsep}. The two datasets are generated in the same way. The first contains density matrices of dimension $9\times 9$ (i.e., $p_A=p_B=3$). The second is composed of density matrices of size $49\times 49$ (i.e., $p_A=p_B=7$).
Each dataset is a collection of pairs of input density matrices and labels indicating whether the corresponding density matrix is separable or entangled, and contains separable~(SEP), PPT entangled (PPT-ENT) and non-PPT (NPPT-ENT) density matrices with 2,000 examples each.
The witnesses obtained via the FW algorithm (Algorithm~\ref{alg:QFW}) that detect the entanglement of the PPT entangled states is also included in the datasets.
For the generation of random density matrices, we used toqito~\cite{russo2021toqito}, an open-source Python library for quantum information that provides many of the same features as the excellent MATLAB toolbox QETLAB~\cite{johnston2016qetlab}.
All the randomly generated density matrices are based on sampling $k$ independent Haar distributed pure states $\{\ket{\psi_i}\}$ on $\mathbb{C}^{p}$~\cite{zyczkowski2011generating}, and considering their uniform mixture
    $\label{eq:random_density}
    \rho = \frac{1}{k}\sum_{i=1}^k \ket{\psi_i} \bra{\psi_i}$.
Previous work on quantum entanglement and random density matrices have shown the existence of a threshold for $k$ above which the generated density matrix $\rho$ is separable with high probability~\cite{aubrun2014random,aubrun2014entanglement}. The matrix $\rho$ is more likely to be entangled with small values of $k$~\cite{ruskai2009bipartite}.

\begin{figure}[t]
\scalebox{0.935}{
\begin{minipage}[t]{0.52\textwidth}
\begin{figure}[H]
    \centering
    \includegraphics[scale=0.12,trim={0 0 0 1.5cm},clip]{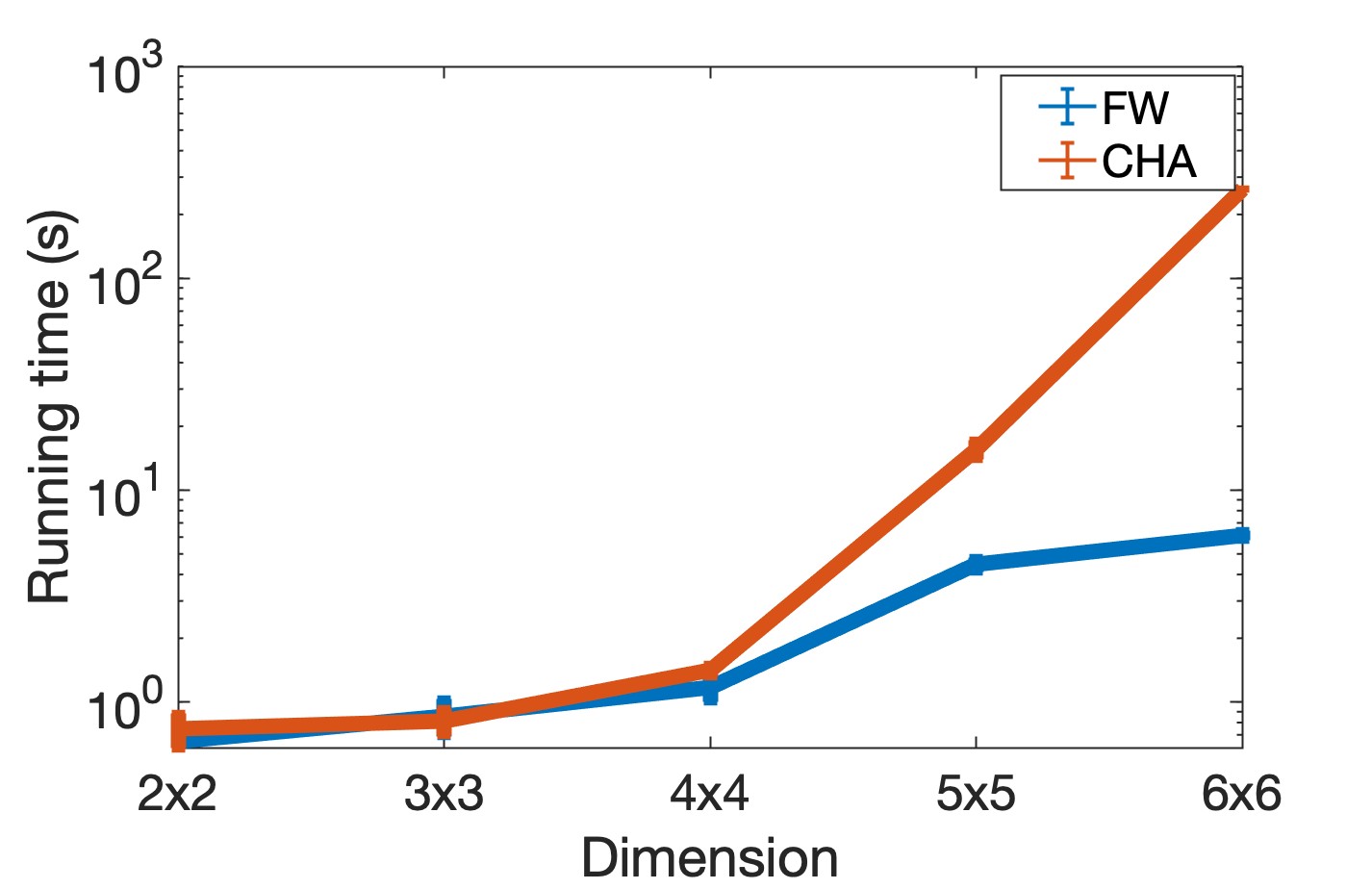}
    \caption{Running time  of 
    CHA~\cite{lu2018separability} and 
    our FW method for density matrices with dimension from $2\times 2$ to $6\times 6$.} 
    \label{fig:cha_vs_fw}
\end{figure}
\end{minipage}
\begin{minipage}[t]{0.55\textwidth}
	\begin{table}[H]
\small
  \caption{Approximation error and running time of Neural Networks (NN)~\cite{girardin2022building} and Frank-Wolfe (FW) method for Werner states of dimension $10\times 10$ parameterized by a scalar $q\in[0,1]$. The results for NN are reported from~\cite{girardin2022building}. The trace distance is the nuclear norm of the difference between the Werner state and its nearest separable approximation.}
  \label{}
  \centering
  \begin{tabular}{lcccc}
    \toprule
     & \multicolumn{3}{c}{Trace distance} & Running time   \\
     Method & q=0 & q=0.5 & q=1 &\\
    \midrule
    NN\footnotemark & 0.09 & 0.045 & 0.5 & 45 min \\
    FW\footnotemark & 0.037 & 0.025 &  0.501 &  58 sec          \\ 
    \bottomrule
  \end{tabular}
    \label{tab:nn_vs_fw}
\end{table}
\end{minipage}
}
\end{figure}
\footnotetext[4]{Trained on a RTX-3080 GPU with 10 GB memory.}
\footnotetext[5]{Timed using a machine with Intel Core i5 CPU, 1.6 GHz with 2 cores and 16 GB RAM.}

\subsection{Running time}

The works in~\cite{lu2018separability}  and \cite{girardin2022building} introduce nearest separable approximations based on convex hull approximation (CHA) and neural network (NN), respectively.
Experiments in the work~\cite{lu2018separability} are limited to density matrices of dimensions $2 \times 2$ and $3 \times 3$, while experiments in~\cite{girardin2022building}  include bipartite quantum states of  dimension up to $10$.
Figure~\ref{fig:cha_vs_fw} shows the running time of CHA and our Frank-Wolfe~(FW) method for density matrices with dimensions varying from $2\times 2$ to $6\times 6$. We set the number of separable states for CHA and the number of iteration for FW to the same value (1500) for a fair comparison. FW is much faster than CHA when the dimension increases  (262 seconds for CHA and 6 seconds for FW with quantum subsystems of dimension 6). CHA is infeasible in practice for large-scale density matrices.

We also compare FW with the NN method using the same setting in~\cite{girardin2022building}. We use Werner states in dimension $10 \times 10$ which are parametrized by a scalar $q\in [0,1]$. Werner states
are separable for $q\leq 0.5$~\cite{girardin2022building}. The approximation error and running time for the two methods are depicted in Table~\ref{tab:nn_vs_fw}. 
FW  achieves comparable error rates to NN (slightly better for $q\leq 0.5$), while being orders of magnitude faster.

\begin{figure}[t]

  \begin{minipage}{0.5\textwidth}
    \centering
    \includegraphics[width=0.7\linewidth]{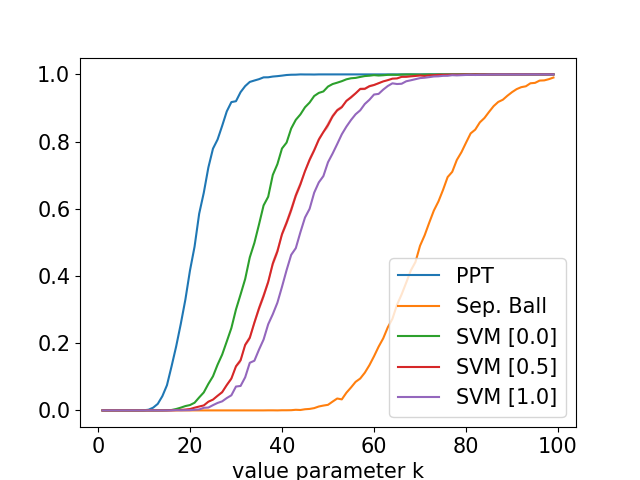}
  \end{minipage}
    \hfill
  \begin{minipage}{0.5\textwidth}
    \centering
    \includegraphics[width=0.7\linewidth]{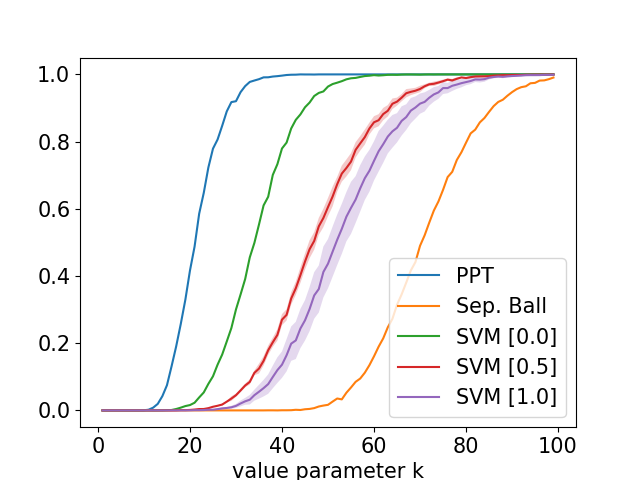}
  \end{minipage}
  
  \caption{Ratio of density matrices predicted as separable in function of the parameter $k$. For SVM, the number in the brackets indicate the portion of entangled PPT states used for learning. Left: Without data augmentation. Right: With data augmentation.}
\label{fig:ratio_dm}
\end{figure}

\subsection{Classification}
We compare the performance of three classifiers trained on the created quantum separability datasets:~(linear and RBF) support vector machines (SVM), neural networks (NN) and bagging convex hull approximation (BCHA~\cite{lu2018separability}). Note that BCHA is specific to the problem of quantum separability and adds a feature to the model by applying CHA  as a preprocess.
All the classifiers are learned with 1,000 examples of the class \enquote{separable} and 1,000 examples of the class \enquote{entangled} in the cases of $3$- and $7$-dimensional qudits ($p_A,p_B=3$ or $7$).
We consider different configurations for the entangled examples depending on the number of entangled PPT states considered for learning.
For data augmentation, we randomly select 10  PPT examples entangled from the datasets and use the process described in Section~\ref{sec:label_augm} to generate a sufficient number of data points. 
The hyperparameters are tuned by five-fold cross-validation.

\paragraph{Learning behaviour with respect to $k$}
In this experiment, test data are generated 
as discussed above with different values of $k$ and a nonlinear SVM with rbf kernel is used. Figure~\ref{fig:ratio_dm} shows the ratio of density matrices predicted as separable in function of the parameter $k$ in the case of $p_A,p_B=3$. We compare the results with those obtained when using the PPT  and the separable ball criteria~\cite{peres1996separability,gurvits2002largest}. Note that the PPT criterion will classify all the entangled PPT states as separable, while the separable ball criterion will detect only separable density matrices contained within a ball centered at the maximally-mixed state, an appropriately-scaled identity matrix~(see Appendix~\ref{sec:appendix_sep} for details). These can be viewed as the two extreme choices for entanglement detection. The obtained results are promising, the percentage of the SVM classifier is in between  these two extreme cases. This means that the SVM should be able to correctly classify more PPT states than the PPT criterion~(those which are entangled) and more separable states than the separable ball criterion.

\begin{figure}[t]
\scalebox{0.9}{
\begin{minipage}[t]{0.5\textwidth}
\begin{table}[H]
    \centering
      \caption{Classification accuracy of different methods on  3-dimensional and 7-dimensional qudit datasets. BCHA becomes
computationally intractable when the dimension of the quantum states grow large.}
    \begin{tabular}{l c c}
    \toprule
     & \multicolumn{2}{c}{Accuracy}   \\
    Method & $3\times 3$ & $7\times 7$ \\
    \midrule
     Linear SVM & $0.539$ ($\pm 0.011$) &  $0.520$ ($\pm 0.165$) \\
    SVM (rbf) & $0.994$ ($\pm 0.003$)  & $0.686$ ($\pm 0.009)$ \\
    Neural Network  & $0.983$ ($\pm 0.002$) & $0.799$ ($\pm 0.071$)\\
    BCHA
    & $0.995$ ($\pm 0.002$) & --- \\
    \bottomrule
    \end{tabular}
    \label{tab:acc_tab_3x3_7x7}
\end{table}
\end{minipage}
\hspace{10pt}
\begin{minipage}[t]{0.5\textwidth}
	\begin{figure}[H]
    \centering
\includegraphics[width=0.8\linewidth]{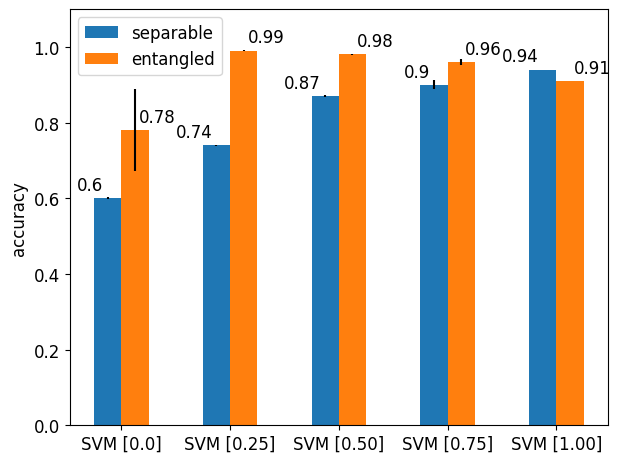}
    \\
    
   \caption{Accuracy of SVM combined with PPT criterion on 7-dimensional qudit dataset.}
    \label{fig:accuracy_7x7}
    \end{figure}
\end{minipage}
}
\end{figure}

\paragraph{Classification Accuracy} We now evaluate the accuracy performance of the classifiers. 
Examples labeled as entangled in the train and test sets are subdivided equally between PPT and NPTT. Accuracy results with the 3-dimensional and 7-dimensional qudit datasets are depicted in Table~\ref{tab:acc_tab_3x3_7x7}. Linear SVM performs very poorly on both datasets. This is expected since the decision boundary between separable and entangled quantum states is not linear. Nonlinear SVM and NN obtain very high accuracy in  the case of 3-dimensional qudits and have similar performance to BCHA. However, their performance drops significantly for quantum states in dimension $7\times 7$.
The task is more challenging due to high-dimensional density matrices.
We also performed experiments on the 3-dimensional qudit dataset using SVM in settings with and without data augmentation, and tried a SVM with 100 training data points instead of 1,000. The performance obtained is lower in this case but remains strong; more than 80\% of classification accuracy. See Appendix~\ref{sec:appendix_exp} for further details.
It is worth noticing that asking a classifier to predict the class NPPT-ENT may not be useful. Indeed, the PPT criterion can be used to label these data correctly.  
So, we combine the predictions of SVM and the PPT criterion on the 7-dimensional qudit dataset. The classification results  are depicted in Figure~\ref{fig:accuracy_7x7}. We get significant gain of performance by combining the decision functions of the two models. This suggests that improvements can arise by cooperating with entanglement detection criteria.
Another observation from Figure~\ref{fig:accuracy_7x7} is that using labeled PPT entangled density matrices in the training process improves the classification accuracy.

\subsection{Limitations} Separable density matrices used during training and test are randomly generated as discussed above. The FW method 
produces separable density matrices.  Since the FW algorithm searches for the nearest separable quantum state of an entangled state, the produced density matrices should be near to the decision frontier between the two classes SEP and PPT-ENT. We create new training and test sets containing also separable density matrices obtained by the FW algorithm~(we call them FW data). The accuracy results are presented in Appendix~\ref{sec:appendix_exp}. 
SVM is unable to predict the correct class label of FW data. Almost all of them are labeled as entangled. When adding FW data in training, the performance is also bad. Our interpretation is that FW data and PPT-ENT data are geometrically close and have different labels, which could make learning harder. We think that metric learning could be an interesting direction to investigate, since it gives the possibility to learn data geometry by taking into account the labels. This is left for future work.

\section{Conclusion}
We studied the problem of quantum separability from a machine learning perspective. We proposed a machine learning pipeline to tackle this problem efficiently. Our approach is based on new computationally efficient procedures for data labelling and augmentation of separable and entangled density matrices. We also provided a large-scale simulated quantum separability dataset to support  reproducible research at the intersection between machine learning and quantum information.


\section*{Acknowledgments}

H.K. would like to thank Guillaume Aubrun for very helpful discussions on random density matrices, and Ronald de Wolf for pointing out work on the robustness of entangled states and for useful comments. This work was supported by funding from the French National Research Agency (QuantML project, grant number ANR-19-CE23-0011).

\begin{appendices}

\section{Bra-Ket Notation}
\label{sec:appendix_braket}

The so called ‘bra-ket’ notation is used to describe the state of a quantum system.
A column vector $\psi$ is represented as \enquote*{ket} $\ket{\psi}$. 
The conjugate transpose (Hermitian transpose) of a ket, a row vector, is denoted by \enquote*{bra} $\bra{\psi} := \ket{\psi}^\dagger$, where $\dagger$ denotes the conjugate transpose.
The inner product of two vectors $\ket{\psi_1}$ and $\ket{\psi_2}$ can be
written in bra-ket notation as $\braket{\psi_1}{\psi_2}$, and their tensor product, can be expressed as $\ket{\psi_1}\ket{\psi_2}$, i.e., $\ket{\psi_1} \otimes \ket{\psi_2}$.

\section{Schmidt Decomposition}
\label{sec:appendix_schmidt}

The different states of a $p$-dimensional quantum system can be represented by a set $\{\ket{0}, \ket{1}, \ldots, \ket{p\!-\!1}\}$ of unit vectors forming an orthonormal basis of a complex Hilbert space~$\Hi$ called the state space of the system. 
A pure quantum state can be represented by a state vector $\ket{\psi}$ in $\mathcal{H}$.
For \textit{bipartite} quantum
systems, i.e., systems composed of two distinct subsystems $A$ and $B$, the Hilbert space $\mathcal{H}$ associated with the composite system is given by  the tensor product $\mathcal{H}_A \otimes \mathcal{H}_B$ of the spaces $\mathcal{H}_A$ and $\mathcal{H}_B$ of the spaces corresponding to each of the subsystems. The dimension $p$ of $\Hi$ is the product of $p_A$ and $p_B$, the dimensions of $\mathcal{H}_A$ and $\mathcal{H}_B$, respectively.

The Schmidt decomposition is a fundamental tool in quantum information theory which plays a central role for entanglement characterization of pure states.

\begin{thm}[{\cite[Theorem 2.7]{nielsen_chuang_2010}}] 
Suppose $\ket{\psi}$ is a pure state of a composite system, $AB$. Then there exist orthonormal states $\{\ket{a_i}\}_{i=1}^{p_A}$ for system $A$, and orthonormal states $\{\ket{b_i}\}_{i=1}^{p_B}$ of system $B$ such that
\begin{equation*}
    \ket{\psi} = \sum_{i=1}^p \lambda_i \ket{a_i} \otimes \ket{b_i},
\end{equation*}
where $\lambda_i$ are non-negative real numbers satisfying  $\sum_i \lambda_i^2=1$ known as Schmidt
coefficients.
\end{thm}
The Schmidt decomposition can be computed numerically by performing a singular value decomposition~\cite[Section 6]{van2000ubiquitous}.

\section{Robustness of PPT Entangled States}
\label{sec:appendix_robust}

Our density matrix data augmentation scheme is based on the notion of robustness of entangled states~(see~\cite{bandyopadhyay2008robustness} and references therein).
This notion quantifies the ability of an entangled quantum states to remain entangled in the presence of noise.
Let $\rho \in \mathcal{S}_{+}^p$ be a PPT entangled density matrix. The paper~\cite{bandyopadhyay2008robustness} investigated how much noise can be added to $\rho$ before it becomes separable, and identified a region $R_\rho$ containing $\rho$ inside of which all the density matrices are also PPT and entangled.
The region $R_\rho$ is characterized analytically:
\begin{equation*}
    R_\rho := \{ (1 - \mu)(\nu \rho + (1 - \nu) \frac{\Id}{p}) + \mu \sigma : \nu \in ]f(W_\rho), 1[, \mu \in [0, g(\nu, W_\rho)[, \sigma \in \mathcal{S}_{+}^p \},
\end{equation*}
where $\sigma$ is a density matrix which corresponds to noise, $\Id$ is the identity matrix, and $f(W_\rho)$ and $g(\nu, W_\rho)$ are scalar-valued function defined as follows
\[
f(W_\rho) = 1 / (1 + \lambda_\rho p),
\]
and
\[
g(\nu, W_\rho) = \min \left[ \frac{1 - \nu}{p - 1 - \nu}, \frac{\mathrm{num}(W_\rho^+)\{\nu(1 + p\lambda_\rho) - 1\}}{p \tr(W_\rho^+) + \mathrm{num}(W_\rho^+)\{\nu(1 + p\lambda_\rho) - 1\}} \right],
\]
where $W_\rho$ is a witness that detects the entanglement of $\rho$, $\lambda_\rho = - \tr(W_\rho \rho)$, $\mathrm{num}(W_\rho^+)$ is the number of positive eigenvalues of $W_\rho$, and $\tr(W_\rho^+)$ is the sum of all positive eigenvalues of $W_\rho$.
Note that all the parameters needed to characterize the region $R_\rho$ can be computed from $\rho$ and its entanglement witness $W_\rho$. 
\begin{thm}[{\cite[Theorem 1]{bandyopadhyay2008robustness}}]
   For any PPT entangled state $\rho$, the region $R_\rho$ contains only PPT entangled states.
\end{thm}
%

\section{Separable Ball Criterion}
\label{sec:appendix_sep}

The separable ball criterion is a simple sufficient condition for separability of
bipartite density matrices~\cite{gurvits2002largest}. 
It detects only separable density matrices contained within a ball centered at the maximally-mixed state, an appropriately-scaled identity matrix.
The size of this ball of separability was provided by the following result.
\begin{corollary}[{\cite[Corollary 3]{gurvits2002largest}}]
    Let $\rho\in\mathcal{S}_+^p$ be a density matrix and ${\alpha}$ the vector of its eigenvalues. $\Id$ denotes the identity matrix and $\mathbbm{1}$ is a vector of all ones. If $\|\rho - \displaystyle\frac{\Id}{p}\|^2 = \|\alpha - \displaystyle\frac{\mathbbm{1}}{p}\|^2 \leq \displaystyle\frac{1}{p(p-1)}:=r^2$, where $\|\cdot\|$ is the Frobenius norm, then $\rho$ is separable. $r$ is the largest such constant.
\end{corollary}
The corollary says that $\rho$ is separable if $\tr(\rho^2)$ is less than or equal to $1/(p-1)$~\cite{gurvits2002largest}.
In quantum information theory, the quantity $\tr(\rho^2)$ is known as the \textit{purity} of a quantum state~\cite{zyczkowski1998volume}. It gives information on how much a state is mixed.

\section{Experimental Details}
\label{sec:appendix_exp}
\subsection{Vector Representation of Density Matrices}
\label{sec:appendix_gellman}

Following~\cite{lu2018separability}, we use the generalized Gell-Mann matrices~(GGM) to obtain a vector representation of density matrices~\cite{bertlmann2008bloch, kimura2003bloch}. GGM are higher–dimensional extensions of the Pauli matrices (for qubits) and the Gell-Mann matrices (for qutrits), which are matrices of dimensions $2\times 2$ and $3\times 3$, respectively.
For a $p$-dimensional space $\Hi$, we have $p^2-1$ GGM defined as follows:
\begin{itemize}
    \item $\displaystyle\frac{p(p-1)}{2}$ symmetric GGM: $S_{i,j} = E_{i,j} + E_{j,i}$, $\forall 1\leq i < j \leq p$,
    \item $\displaystyle\frac{p(p-1)}{2}$ antisymmetric GGM: $A_{i,j} = -i E_{i,j} + iE_{j,i}$, $\forall 1\leq i < j \leq p$,
    \item $(p-1)$ diagonal GGM: $D_k = \displaystyle\sqrt{\frac{2}{k(k+1)}} \left( \sum_{i=1}^k E_{i,i} - k E_{k+1,k+1}\right)$, $\forall 1 \leq k < p$,
\end{itemize}
where $E_{i,j}$, $\forall 1\leq i,j \leq p$, is the $p\times p $ matrix whose element $(i,j)$ is 1 and the other elements are 0.

GGM are orthogonal and form a basis~\cite{bertlmann2008bloch}. Let $\{G_l\} = \{S_{i,j}\} \cup \{A_{i,j}\}  \cup \{D_k\}$ be the set of the $p^2-1$ GGM. The vector representation of a density matrix $\rho\in \mathcal{S}_+^p$, a.k.a Bloch vector expansion, is obtained from the following expression:
\begin{equation*}
    \rho = \frac{1}{p}\mathbb{I} + \sum_{l=1}^{{p}^2-1} \beta_l G_l,
\end{equation*}
where $\beta_l = \tr(G_l\rho)$ and the vector $\beta = (\beta_1,\ldots, \beta_{{p}^2-1} )$ is the Bloch vector.
Note that $\beta_l$, $\forall l$, is a real number and then $\beta\in\mathbb{R}^{{p}^2-1}$.

\subsection{SVM and neural network classifiers}
\label{sec:appendix_svm}

We train a nonlinear SVM with polynomial and Gaussian kernels and use the scikit-learn SVM implementation~\cite{pedregosa2011scikit}.
The hyperparameters of the SVM and the kernels were determined by 5-fold cross validation  without using the test data.
Code and trained models are available at \url{https://gitlab.lis-lab.fr/balthazar.casale/ML-Quant-Sep}.

For neural networks,  we used two hidden layers with $2d$ neurons in the first layer and $d$ neurons in the second, where $d$ is the dimension of the data.   We also used Relu activation, batch normalization  and Adam with a learning rate of $0.001$.

\subsection{Additional results}
\label{sec:appendix_additional}

Accuracy results with the 3-dimensional qudit dataset 
with different ratios of PPT data are depicted in Table~\ref{tab:accuracy_3x3}. SVM obtains very high accuracy with and without data augmentation. Note that we use the same number of data in the two configurations in order to validate the proposed density matrix data augmentation scheme.

Since SVM achieves good performance on the $3\times 3$ quantum dataset when trained with 1000 examples per class, we also performed experiments with an SVM trained using 100 examples. The results are shwon in Table~\ref{test_score_3x3}. It is interesting to see that our data augmentation strategy helps to improve the prediction accuracy.

FW data are generated using the proposed Frank-Wolfe method for approximating the nearest separable density matrix. These data are near to the decision frontier between the two classes SEP and PPT-ENT. The accuracy results of SVM when using this data are presented in Table~\ref{tab:fw_data}.  The presence of FW data largely reduces the performance of SVM.

\begin{table}[t]
  \caption{Classification accuracy of SVM on 3-dimensional qudit dataset with 1,000 examples per class.}
  \label{test_score_3x3}
  \centering
  \begin{tabular}{llll}
    \toprule
    PPT ratio & score SEP & score PPT-ENT & score NPPT-ENT \\
    \midrule
    $0\%$ & $0.994$ ($\pm 0.000$) & $0.969$ ($\pm 0.002)$ & $0.997$ ($\pm 0.000$) \\
    \midrule
    \multicolumn{4}{c}{without augmentation}                   \\
    \midrule
    $25\%$ & $0.993$ ($\pm 0.001$) & $0.984$ ($\pm 0.003$) & $0.998$ ($\pm 0.001$) \\
    $50\%$ & $0.994$ ($\pm 0.001$) & $0.991$ ($\pm 0.001$) & $0.998$ ($\pm 0.001$) \\
    $75\%$ &  $0.994$ ($\pm 0.001$) & $0.994$ ($\pm 0.001$) & $1.000$ ($\pm 0.000$)  \\
    $100\%$ & $0.994$ ($\pm 0.000$) & $0.994$ ($\pm 0.000$) & $1.000$ ($\pm 0.000$)  \\
    \midrule
    \multicolumn{4}{c}{with augmentation} \\
    \midrule
    $25\%$ & $0.993$ ($\pm 0.001$) & $0.990$ ($\pm 0.004$) & $0.998$ ($\pm 0.001$) \\
    $50\%$ & $0.994$ ($\pm 0.001$) & $0.998$ ($\pm 0.001)$ & $1.000$ ($\pm 0.000$) \\
    $75\%$ & $0.991$ ($\pm 0.002$) & $0.999$ ($\pm 0.001$) & $1.000$ ($\pm 0.000$) \\
    $100\%$ & $0.992$ ($\pm 0.001$) & $0.999$ ($\pm 0.000$) & $1.000$ ($\pm 0.000$) \\
    \bottomrule \\
  \end{tabular}
    \label{tab:accuracy_3x3}

\end{table}

\bigskip

\begin{table}[t]
\small
  \caption{Classification accuracy of SVM on 3-dimensional qudit dataset with 100 examples per class.}
  \label{test_score_3x3}
  \centering
  \begin{tabular}{llll}
    \toprule
    PPT ratio & score SEP & score PPT-ENT & score NPPT-ENT \\
    \midrule
    $0\%$ &  $0.922$ ($\pm 0.005$) & $0.648$ ($\pm 0.032)$ & $0.822$ ($\pm 0.032$) \\
    \midrule
    \multicolumn{4}{c}{without augmentation}                   \\
    \midrule
    $25\%$ &$0.934$ ($\pm 0.025$) & $0.668$ ($\pm 0.039$) & $0.817$ ($\pm 0.044$)     \\
    $50\%$ &  $0.914$ ($\pm 0.016$) & $0.676$ ($\pm 0.035$) & $0.822$ ($\pm 0.016$)     \\
    $75\%$ & $0.911$ ($\pm 0.014$) & $0.742$ ($\pm 0.047$) & $0.877$ ($\pm 0.038$)     \\
    $100\%$ &$0.915$ ($\pm 0.007$) & $0.693$ ($\pm 0.057$) & $0.824$ ($\pm 0.049$)    \\
    \midrule
    \multicolumn{4}{c}{with augmentation} \\
    \midrule
    $25\%$ & $0.918$ ($\pm 0.006$) & $0.687$ ($\pm 0.042$) & $0.814$ ($\pm 0.033$)     \\
    $50\%$ &  $0.906$ ($\pm 0.015$) & $0.737$ ($\pm 0.051)$ & $0.842$ ($\pm 0.040$)     \\
    $75\%$ &$0.908$ ($\pm 0.009$) & $0.786$ ($\pm 0.026$) & $0.891$ ($\pm 0.018$)    \\
    $100\%$ & $0.901$ ($\pm 0.011$) & $0.821$ ($\pm 0.037$) & $0.911$ ($\pm 0.031$)    \\
   \bottomrule \\
  \end{tabular}
  \label{tab:accuracy_3x3_complete}
\end{table}

    \begin{table}[H]
		\caption{Classification accuracy of SVM when using separable density matrices generated by the FW method~(Algorithm~\ref{alg:QFW}) during train and test.}
		\label{frank-wolfe effect}
		\centering
		\begin{tabular}{llll}
			\toprule
			SEP & FW & PPT-ENT & NPPT-ENT \\
			\midrule
			\multicolumn{4}{c}{without FW data} \\
			\midrule
			$0.99$ ($\pm 0.0$) & $0.01$ ($\pm 0.0$) & $0.99$ ($\pm 0.04$) & $0.99$ ($\pm 0.01$) \\
			\midrule
			\multicolumn{4}{c}{with FW data} \\
			\midrule
			$0.97$ ($\pm 0.01$) & $0.21$ ($\pm 0.02$) & $0.76$ ($\pm 0.03$) & $0.88$ ($\pm 0.02$) \\
			\bottomrule
		\end{tabular}
		\label{tab:fw_data}
	\end{table}

\end{appendices}

\bibliographystyle{plain}
\bibliography{ml_quantum_sep}

\end{document}